\documentclass{kapproc} 
\usepackage{procps} 
\usepackage[dvips]{graphicx}
\upperandlowercase
\kluwerbib

\begin{document}

\articletitle{Pattern Speeds of Spiral Galaxies using the Tremaine-Weinberg Method}
\author{S. E. Meidt\altaffilmark{1} and R. J. Rand\altaffilmark{1}}

\altaffiltext{1}{University of New Mexico}

\section*{Introduction}
Long-lasting spiral structure in galaxies is established when density waves form a pattern that rotates at a constant 
angular frequency $\Omega_p$, the pattern speed.  This quantity, critical to understanding spiral structure, 
galaxy evolution, and the connection, if any, between bars and spirals in galaxies, is not directly measurable. 
Most determinations of pattern speeds have relied on identifying morphological features at certain radii as 
resonances, but this depends significantly on knowledge of the behavior of stars and gas at these resonances. 
The Tremaine-Weinberg (TW) method allows for the determination of pattern speeds based on kinematic observations 
in a way completely independent of resonance identification, or theories of spiral structure, relying on the 
identification of a tracer that obeys the continuity equation as it orbits through the pattern. \\
\indent
This work expands on previous studies of the TW method applied to galaxies with 
molecular- or HI-dominated ISMs.  In galaxies where the ISM is found to be dominated by one phase, it is 
assumed that this component satisfies the continuity requirement; for galaxies with $\rm{H}_2$ dominated ISMs, for instance, 
the TW method has successfully been applied using $^{12}$CO(1-0) emission as a tracer for the molecular gas surface 
density (\cite{randwallin} 2004).  
\section{The Tremaine-Weinberg Method}
If the component obeys continuity as it orbits in the spiral/bar potential, \cite{tw} show that
\begin{equation}
\Omega_p \sin{i}=\frac{\int\limits_{-\infty}^\infty {\Sigma(X,Y) V_{\parallel}}dX}{\int\limits_{-\infty}^\infty {\Sigma(X,Y) X dX}},
\end{equation}
\noindent where $(X,Y)$ are the sky coordinates $\parallel$ and $\perp$ to the major axis; $\Sigma(X,Y)$ is the surface 
density of the component; $V_\parallel$ is the line-of-sight velocity; and $i$ is the inclination angle of the galaxy.\\
\indent  
If the intensity of the component's emission traces the surface density, then, at a given $Y$ (cf. \cite{merrifield}), 
\begin{equation}
\Omega_p =\frac{1}{\sin{i}}\frac{\int\limits_{-\infty}^\infty {I(X) V_{\parallel}}dX}{\int\limits_{-\infty}^\infty {I(X) X dX}}\frac{\int\limits_{-\infty}^\infty {I(X)dX}}{\int\limits_{-\infty}^\infty {I(X) dX}},
\end{equation}
\noindent and $\Omega_p$, modulo $\sin{i}$, can be determined from total intensity and velocity field maps by measuring 
intensity-weighted averages of observed position <x> and velocity <v> along lines of constant $Y$
(parallel to the major axis).  The slope in a plot of <v> vs. <x> for such "apertures" is the pattern speed. 

\section{Data and Analysis}
Presently, to determine the spiral density wave pattern speed, we use high-resolution CO data cubes from the BIMA 
Survey of Nearby Galaxies (SONG).  The BIMA SONG 
observations include single-dish data, thereby sampling all spatial frequencies to support the continuity requirement.  
Pattern speed determinations for three of the galaxies in this work are presented based on CO emission data cubes 
alone; we cite the results of \cite{wb} wherein radial profiles of NGC 5055, 5033, and 5457 (M101) 
indicate that these galaxies are molecule-dominated within the region of CO detection.  \\
\indent In this work, a constant conversion factor $X$=2$\times$10$^{20}$ mol cm$^{-2}$(K km s$^{-1}$)$^{-1}$ between 
CO intensity and $\rm{H}_2$ column density is assumed. As demonstrated in \cite{zimmer}, the choice of conversion factor 
has little effect on the TW calculation.  The choice, however, clearly influences assertions of molecular 
dominance, particularly with regard to an important suspected variation of $X$ with metallicity, which is supported by 
empirical evidence for an inverse
correlation (e.g. \cite{boselli}). For the galaxies studied here, NGC 5457, 5055 and 5033, 
metallicity estimates are subsolar (as compiled by \cite{metal}), indicating 
that, if anything, the conversion factor underestimates the amount of $\rm{H}_2$ present.\\
\indent Inclination and position angles (PA's) used in the TW
calculation are determined by fitting tilted rings to the velocity field as performed by the GIPSY task ROTCUR.  
As found in \cite{debattista} and \cite{randwallin} (2004), the scatter in the slope of the <v> vs. <x> plot worsens for PA's 
further from the nominal value.   Error bars reported for the pattern speed of each galaxy 
account for PA uncertainty and the scatter in the <v> vs. <x> plot for nominal PA.   

\section{Results}

\begin{table}
\begin{center}
\begin{tabular}{|r|c|c|c|c|c|c|c|}
\hline
&Morphological&D&i&PA &$V_{sys}$&$\Omega_p$\\
&Type&(Mpc)&($^o$)&($^o$)&(km/s)&(km/s/kpc)\\
\hline
NGC 5457\hspace{1 pc}&SAB(rs)cd&7.4&21&42 +4/-2&257&48 +7/-6\\
\hline
NGC 5055&SA(rs)bc&7.2&57&99 $ \pm$1&506&67 $ \pm$6\\
\hline
NGC 5033&SA(s)c&11.8&68&352 $ \pm$2&880&93 +18/-39\\
\hline
\end{tabular}
\caption{Properties of sample galaxies.}
\end{center}
\end{table}

\begin{figure}[b]
\hspace{.25 cm}\includegraphics[width=1.75 in]{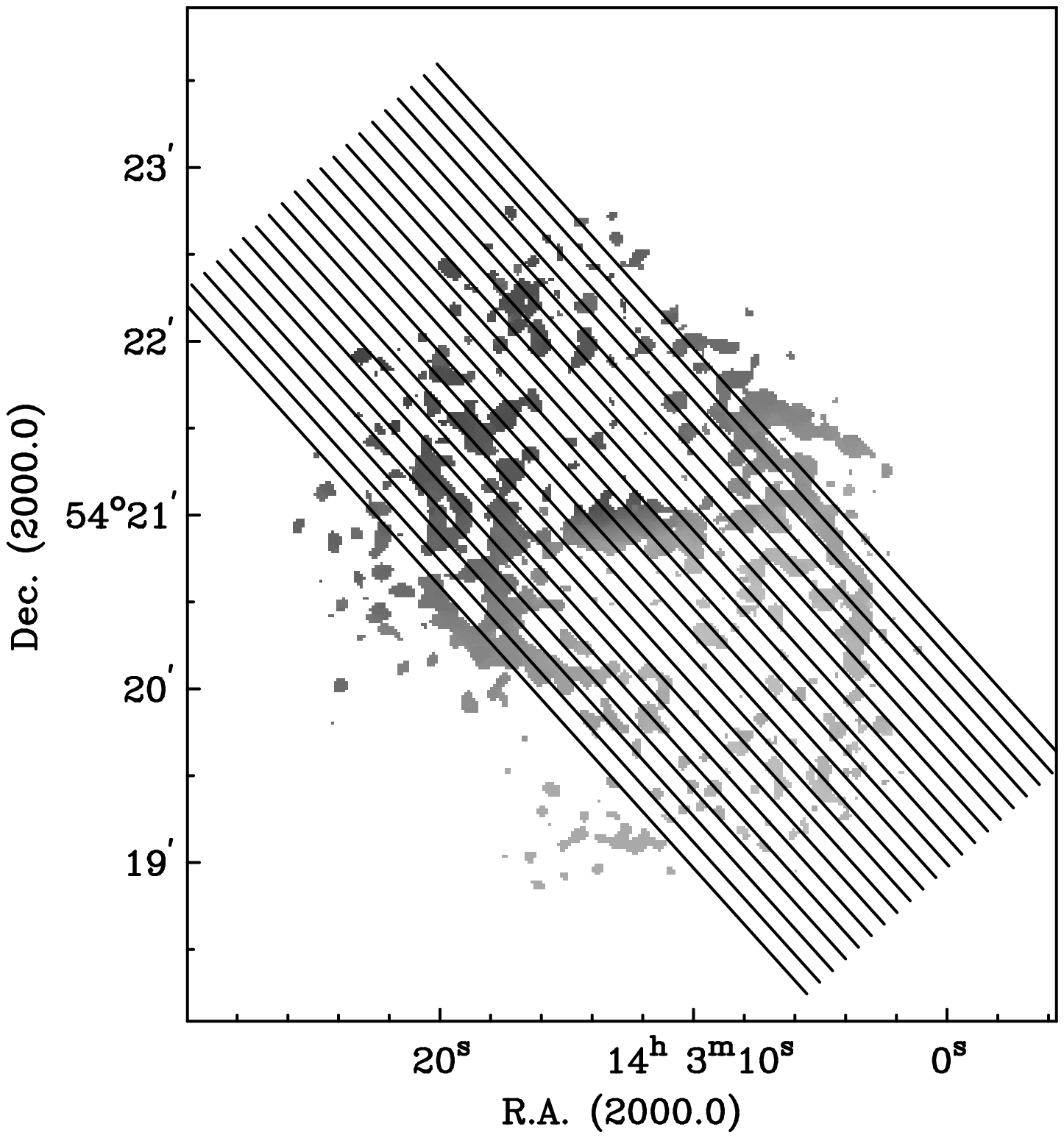}\hspace{1.55 cm}\includegraphics[width=2.25 in]{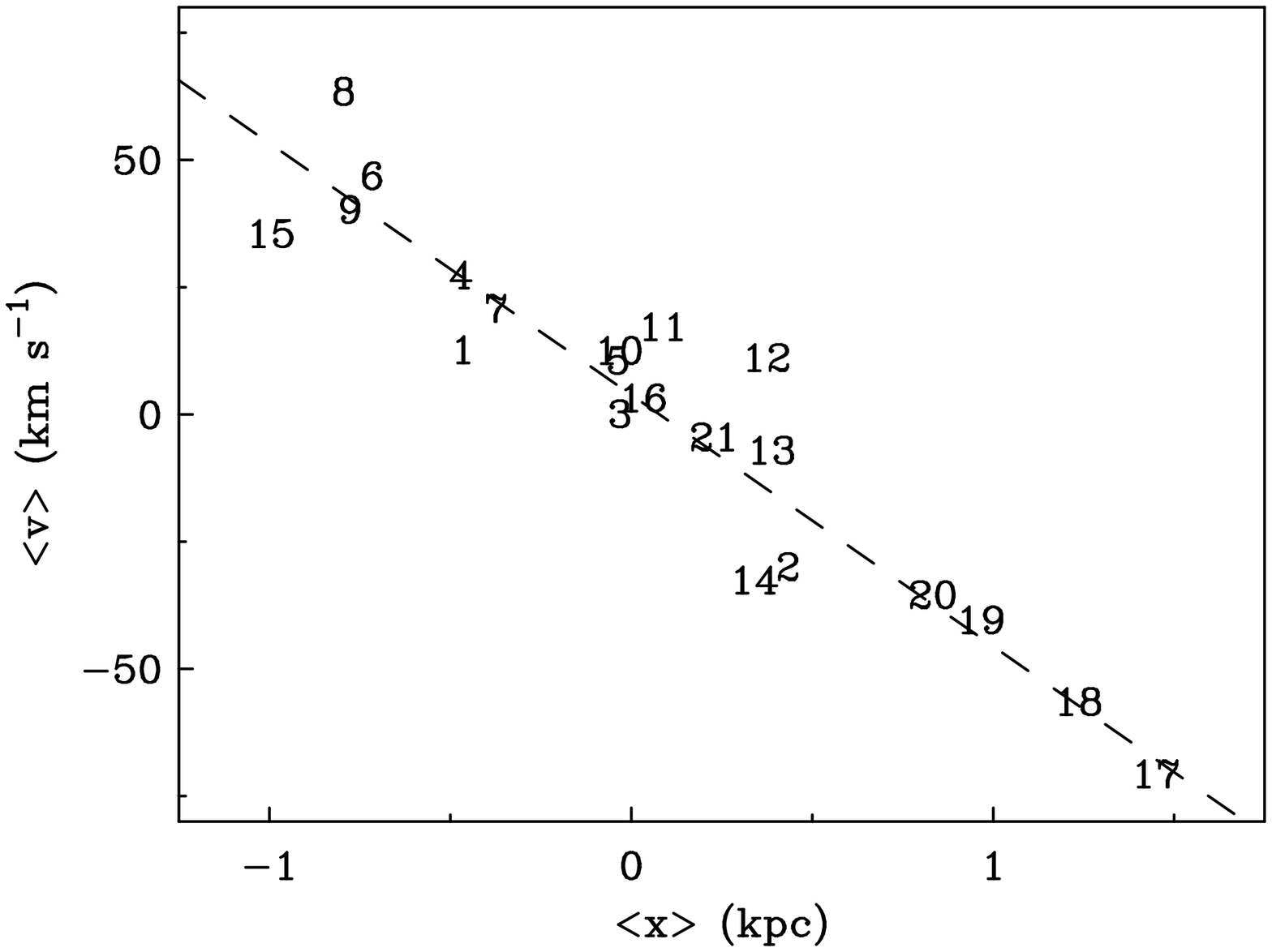}
\sidebyside{\letteredcaption{a}{First moment map of the BIMA CO data for NGC 5457 with apertures used in the
TW calculation spaced at the resolution of the map shown overlayed as solid lines, with the 
westernmost being aperture 1.}}
{\letteredcaption{b}{Plot of <v> vs. <x> for the 21 apertures shown in 
fig.1a. The dashed line is the best-fit straight line to all apertures.  The corresponding value of $\Omega_p$ is
48 +7/-6 km s$^{-1}$ kpc$^{-1}$.}}
\end{figure}

Given that NGC 5457, 5055 and 5033 appear to be dominated by molecular gas in their inner disks, 
we can be reasonably confident that the continuity requirement of the TW method is satisfied.  Table 1 lists the 
properties and results of the TW calculation for each of the galaxies in our sample.  
Figures 1a and b display an example of the line-of-sight apertures used in the TW calculation for NGC 5457 
and the corresponding <v> vs. <x> plot, respectively.
\section{Conclusions and Future Work}
The Tremaine-Weinberg method has long been applied with much success to bars in galaxies.  The relatively 
recent applications of the method to spiral patterns, too, are compelling: pattern speeds for M51 and M83, 
for example (\cite{randwallin} 2004), were found to be in good agreement with those determined from resonances.  
Generally, however, it is unclear how reliable spiral pattern speed determinations with the TW method are, 
considering that the converging and diverging flows responsible for the pattern are typically weaker in a 
spiral (e.g. \cite{robertstewart}) than in a bar; the method, based on the continuity equation, relies on 
the detection of such non-axisymmetric, or "streaming", motions that trace spiral arm surface density enhancements 
in order to calculate a pattern speed.  It is easy to show, in the case of purely axisymmetric azimuthal motion, 
that the method returns a weighted average of the angular rotation frequency.  This makes sense: in such a case, 
only a material pattern can be sustained. \\ 
\indent To understand whether streaming motions are significantly contributing 
to the TW analysis, we have repeated the procedure using axisymmetric velocity fields built from the output of ROTCUR.
We generally find little difference relative to the result for the original velocity field (shown here for the case 
of NGC 5457 in a plot of <v>/<x> vs. aperture number in Figure 2).  
\begin{figure}[t]
\hspace{0.5 cm}\includegraphics[width=2.25 in]{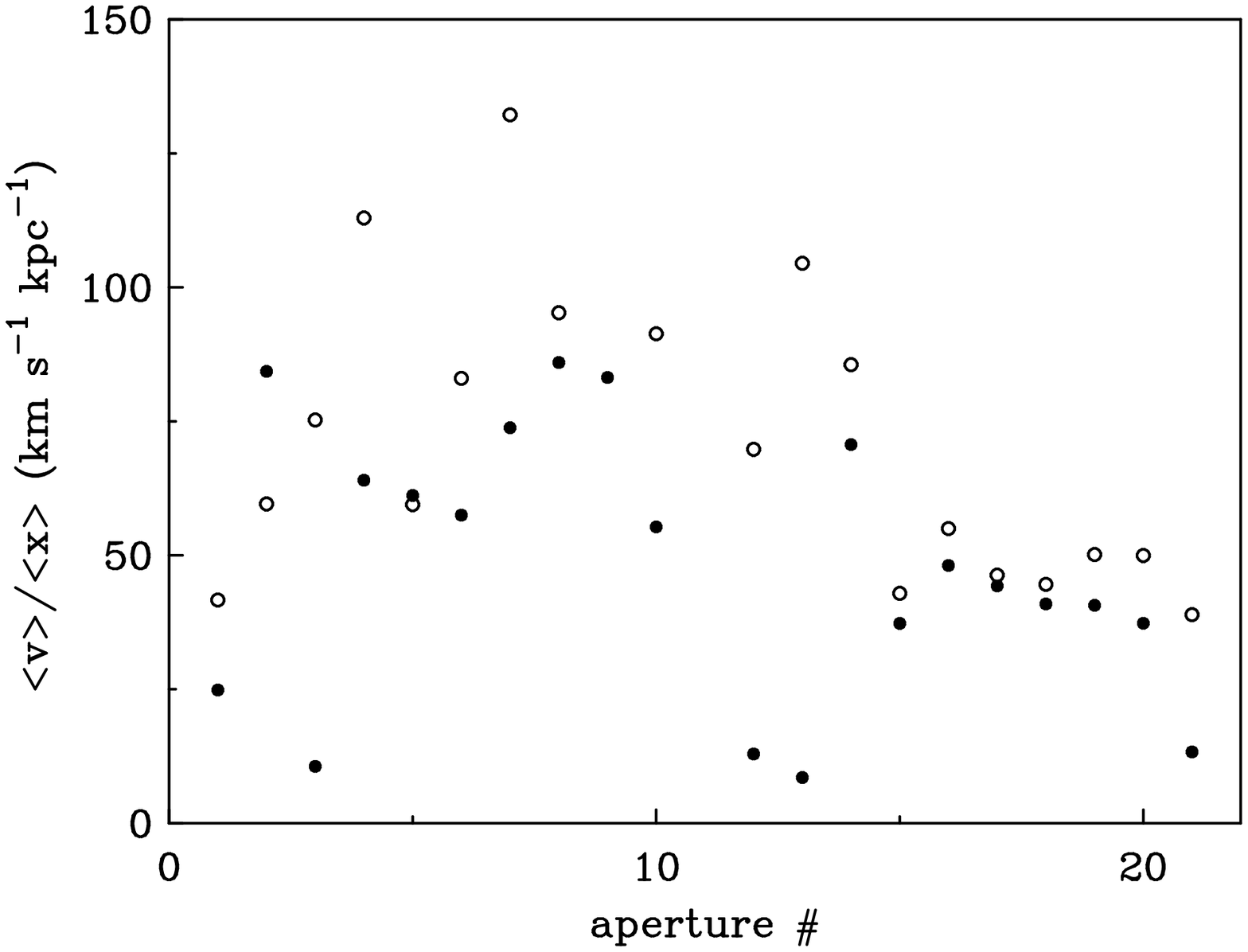}
\narrowcaption{Plot of <v>/<x> vs. aperture number for NGC 5457 with filled and open circles representing values from 
the TW calculation using the original and axisymmetric velocity fields, respectively.\newline\newline\newline}
\end{figure}
While this may indicate a limitation of the method, it may alternatively indicate that spiral patterns are in fact 
close to material patterns.  With M. Merrifield, we are pursuing the latter possibility, while with P. Rautiainen 
we are studying applications of the TW method on simulated spirals with varying pattern speeds and spiral strengths 
in order to assess the reliability of the method on spiral patterns.  

\end{document}